\documentclass[american,prl,notitlepage,reprint,superscriptaddress]{revtex4-1}
\pdfoutput=1
\usepackage[T1]{fontenc}
\usepackage[latin9]{inputenc}
\usepackage{babel}
\usepackage{bm}
\usepackage{amsmath}
\usepackage{amsthm}
\usepackage{amssymb}
\usepackage{graphicx}
\usepackage{physics}
\usepackage[colorlinks=true,urlcolor=blue,citecolor=blue,linkcolor=blue]{hyperref}
\PassOptionsToPackage{linktocpage=true}{hyperref}

\makeatletter
\theoremstyle{remark}
\newtheorem*{rem*}{\protect\remarkname}

\makeatother

\providecommand{\remarkname}{Remark}

\begin{document}

\title{Exactly solvable Majorana-Anderson impurity models}

\author{G. Shankar}
\affiliation{Department of Physics, University of Alberta, Edmonton, Alberta T6G 2E1, Canada}
\author{J. Maciejko}
\affiliation{Department of Physics, University of Alberta, Edmonton, Alberta T6G 2E1, Canada}
\affiliation{Theoretical Physics Institute, University of Alberta, Edmonton, Alberta T6G 2E1, Canada}
\affiliation{Canadian Institute for Advanced Research, Toronto, Ontario M5G 1Z8, Canada}

\date\today
\begin{abstract}
Motivated by recent experimental progress in the realization of hybrid structures with a topologically superconducting nanowire coupled to a quantum dot, viewed through the lens of the emerging field of correlated Majorana fermions, we introduce a class of interacting Majorana-Anderson impurity models which admit an exact solution for a wide range of parameters, including on-site repulsive interactions of arbitrary strength. The model is solved by mapping it via the $\mathbb{Z}_2$ slave-spin method to a noninteracting resonant level model for auxiliary Majorana degrees of freedom. The resulting gauge constraint is eliminated by exploiting the transformation properties of the Hamiltonian under a special local particle-hole transformation. For a spin-polarized Kitaev chain coupled to a quantum dot, we obtain exact expressions for the dot spectral functions at both zero and finite temperature. We study how the interaction strength and localization length of the end Majorana zero mode affect physical properties of the dot, such as quasiparticle weight, double occupancy, and odd-frequency pairing correlations, as well as the local electronic density of states in the superconducting chain.
\end{abstract}
\maketitle

\emph{Introduction.}---The discovery of topological phases of quantum matter has led to a paradigm shift in condensed matter physics. The simplest such topological phase, the one-dimensional (1D) topological superconductor (SC)~\cite{Kitaev2001}, hosts localized Majorana zero modes (MZMs) at its ends which can form a topological qubit immune to decoherence, with exciting prospects for quantum computation~\cite{kitaev2003,nayak2008}. Strong evidence suggests MZMs have been observed in experiments on proximitized semiconductor nanowires~\cite{mourik2012} and ferromagnetic chains~\cite{nadj-perge2014}, following specific theoretical proposals~\cite{lutchyn2010,oreg2010,nadj-perge2014}.

On the theoretical front, a new direction has emerged which explores the interplay of pure MZM physics, well understood from single-particle quantum mechanics, and electronic correlations~\cite{RahmaniFranzReview}. Recently studied lattice models of interacting MZMs such as the Majorana-Hubbard~\cite{rahmani2015,*rahmani2015b,affleck2017,
li2018,wamer2018,rahmani2019,hayata2017} and Majorana-Falicov-Kimball~\cite{Prosko2017,li2019} models may be relevant to describe Abrikosov vortex lattices in 2D topological SCs~\cite{chiu2015}, where each vortex hosts an unpaired MZM~\cite{read2000,fu2008}. 
Motivated by transport experiments on proximitized nanowires, another avenue of research has explored interacting Anderson-type quantum impurity models involving small numbers of MZMs coupled to dissipative baths, some of which are predicted to exhibit exotic Kondo effects~\cite{beri2012,altland2013,herviou2016}. A geometry of particular interest, that of an end MZM tunnel-coupled to a quantum dot (QD), is now experimentally accessible~\cite{Deng2016} and argued to directly probe the nonlocality of MZMs~\cite{leijnse2011,liu2011,lee2013,Cheng2014,liu2015,
clarke2017,prada2017}. Existing theoretical studies of this problem have largely relied on mean-field approximations~\cite{Cheng2014,liu2015,prada2017} or numerical methods~\cite{lee2013,Cheng2014} to treat correlation effects in the corresponding Anderson model~\footnote{In Ref.~\cite{Cheng2014} an exact solution was presented, but for an effective Kondo model to which the Anderson model reduces in the limit of infinite on-site repulsion $U\rightarrow\infty$ on the QD, in which charge fluctuations are completely frozen out. By contrast, our exactly solvable model keeps $U$ finite and retains the degrees of freedom associated with charge fluctuations}.  Such studies also typically model the MZM as a unique on-site Majorana operator, whereas the MZM localization length is generically finite, as known from both theory~\cite{Kitaev2001} and experiment~\cite{albrecht2016}. In this work, we introduce a class of Majorana-Anderson impurity (MAI) models which admit an exact solution regardless of interaction strength and the degree of MZM localization.

\emph{Majorana-Anderson impurity models and exact solvability.}---We
consider a class of models described
by a lattice Hamiltonian of the form $H\!=\!H_{C}\!+\!H_{A}\!+\!H_{\text{hyb}}$, 
where $H_{C}$ describes either a host material or leads that couple
to the QD, and is quadratic in spinless fermion
operators $c_j$, $c_j^\dag$ where $j$ is a site index. The QD is modeled as an Anderson impurity,
\begin{equation}
H_{A}\!=\!U\prod_\sigma\left(2n_{d\sigma}\!-\!1\right)\!+\!\frac{\epsilon}{2}\left(n_{d\uparrow}\!+\!n_{d\downarrow}\!-\!1\right)\!-\!\frac{h}{2}\left(n_{d\uparrow}\!-\!n_{d\downarrow}\right),\label{eq:pHA}
\end{equation}
where $n_{d\sigma}\!=\!d_{\sigma}^{\dagger}d_{\sigma}$ is the number
operator for fermions of spin $\sigma\!\in\!\{\uparrow,\downarrow\}$ on the impurity. $U$ describes on-site Coulomb repulsion,
$h$ is a Zeeman field, and $\epsilon$ is a shift in the chemical
potentials of the impurity fermions. The hybridization between the
host and QD is
\begin{equation}
H_{\text{hyb}}=-i\sum_{j}V_{j}(c_{j}+c_{j}^{\dagger})(d_{\uparrow}+d_{\uparrow}^{\dagger}),\label{eq:pHhyb}
\end{equation}
which allows for the possibility of spatially extended hybridization
(strength $V_{j}$) between the QD and host. This form of Majorana
hybridization arises naturally if the host supports a localized MZM that is in proximity to an impurity.
As MZMs arise in effectively spin-polarized SCs, it is
reasonable to expect that only one impurity spin species will hybridize~\cite{hoffman2017,lee2013,Cheng2014,liu2015}. The number $n_{d\downarrow}$ of spin-$\downarrow$ fermions being thus conserved, the problem studied here can be thought of as a Majorana version of the X-ray edge problem~\cite{mahan1967,nozieres1969}. By contrast with the classic Nozi\`eres-De~Dominicis solution of the original problem~\cite{nozieres1969}, which is restricted to asymptotically low frequencies, here we find an exact solution for the impurity spectral functions at all frequencies.

The key ingredient in constructing an exact solution for the MAI model is the $\mathbb{Z}_{2}$ slave-spin method pioneered by R\"{u}egg \emph{et
al.}~\cite{Ruegg2010,huber2009} and since employed in a variety of contexts ranging from non-Fermi liquids~\cite{nandkishore2012} to fractionalized topological phases~\cite{ruegg2012,maciejko2013,maciejko2014,prychynenko2016} and the Mott transition in infinite dimensions~\cite{zitko2015}. Following Ref.~\cite{Ruegg2010}, we fractionalize the physical impurity fermions
into an Ising slave pseudospin and slave fermions as $d_{\sigma}^{\left(\dagger\right)}\!=\!\mu^{x}f_{\sigma}^{\left(\dagger\right)}$,
where $\sigma\!\in\!\{\uparrow,\downarrow\}$ is the spin projection (along $\hat{z}$) of the physical $(d)$ and slave ($f$) fermions, and $\{\mu^{x},\mu^{y},\mu^{z}\}$ are Pauli matrices that describe the auxiliary slave pseudospin. Physical states in the enlarged Hilbert space satisfy the gauge constraint
\begin{equation}
\mu^{z}=2\left(n_{f}-1\right)^{2}-1,\label{eq:ctr}
\end{equation}
 where $n_{f}\!=\!f_{\uparrow}^{\dagger}f_{\uparrow}\!+\!f_{\downarrow}^{\dagger}f_{\downarrow}$
is the total number of slave fermions. The constraint can be
used to construct a projector 
\begin{equation}
\mathbb{P}=\frac{1}{2}\left[1+\left(-1\right)^{n_{f}}\mu^{z}\right],\label{eq:projS}
\end{equation}
 that projects onto the physical subspace. The
slave-spin (SS) representation of $H$ in the physical subspace is then 
\begin{multline}
H_{SS}=H_{C}-i\sum_{j}V_{j}(c_{j}+c_{j}^{\dagger})(f_{\uparrow}+f_{\uparrow}^{\dagger})\mu^{x}+U\mu^{z}\\
+\frac{1}{2}\left[\epsilon+h+(\epsilon-h)\mu^{z}\right]\left(n_{f\downarrow}-1/2\right),\label{eq:ssH1}
\end{multline}
where the constraint equation has been used to rewrite the interaction,
chemical potential, and Zeeman terms \cite{Guerci2019}. Defining
new Majorana operators $\Gamma_{\uparrow}^{\alpha}\!=\!\mu^{\alpha}(f_{\uparrow}+f_{\uparrow}^{\dagger})$
where $\alpha\!\in\!\{x,y,z\}$, and using $\mu^{z}\!=\!-i\mu^{x}\mu^{y}\!=\!-i\Gamma_{\uparrow}^{x}\Gamma_{\uparrow}^{y},$
the slave-spin Hamiltonian can be written entirely in terms of fermion operators as
\begin{multline}
H_{SS}=H_{C}-i\sum_{j}V_{j}(c_{j}+c_{j}^{\dagger})\Gamma_{\uparrow}^{x}-iU\Gamma_{\uparrow}^{x}\Gamma_{\uparrow}^{y}\\
+\frac{1}{2}\left[\epsilon+h-i(\epsilon-h)\Gamma_{\uparrow}^{x}\Gamma_{\uparrow}^{y}\right]\left(n_{f\downarrow}-1/2\right).\label{eq:ssH2}
\end{multline}
For $\epsilon\!=\!h$, this model is bilinear in fermions and thus exactly
solvable. Henceforth, we set $\epsilon\!=\!h$, and consider deviations
from this exactly solvable limit later. In an experimental situation we expect $\epsilon$ and $h$ to be tunable via gate potentials and applied magnetic fields, respectively.

The physical partition function for MAI models can be computed in the SS representation without constraint, even away from the exactly solvable point. The proof is similar
to those for other such constraint-free models studied using the
$\mathbb{Z}_{2}$ slave-spin method~\cite{zitko2015,Guerci2017,Prosko2017,Guerci2019}.
Defining a particle-hole transformation $\mathcal{D}_{\uparrow}$
that acts only on $d_{\uparrow}$ as $\mathcal{D}_{\uparrow}d_{\uparrow}\mathcal{D}_{\uparrow}^{-1}\!=\!d_{\uparrow}^{\dagger}$,
Eqs.~(\ref{eq:pHA})-(\ref{eq:pHhyb}) yield $\mathcal{D}_{\uparrow}H(V,U,\epsilon,h)\mathcal{D}_{\uparrow}^{-1}\!=\!H(V,-U,h,\epsilon)$. Since the partition function is invariant under similarity
transformations of the Hamiltonian, $\mathcal{Z}(V,U,\epsilon,h)\!=\!\mathcal{Z}(V,-U,h,\epsilon)$. This transformation is implemented in the SS representation (on $H_{SS}$) by $\mu^{x}$. Using cyclicity of the trace and the relation $\mu^x\mathbb{P}\mu^x\!=\!1\!-\!\mathbb{P}$, it is easy to show that $\mathcal{Z}\!=\!\mathcal{Z}_{SS}/2$. Similarly, it
can also be shown that correlation functions of operators that commute
with $\mathcal{D_{\uparrow}}$ are calculable without constraint~\cite{SuppMat}. However, for MAI models, it
is possible to exactly implement the constraint and compute \emph{all
}correlation functions in the SS representation. To see this, note
that the projector $\mathbb{P}$ admits a fermion representation,
\begin{equation}
\mathbb{P}=i\Gamma_{\uparrow}^{z}\text{\ensuremath{\gamma'_{f\uparrow}}}(f_{\downarrow}^{\dagger}f_{\downarrow}-1/2)+1/2,\label{eq:projF}
\end{equation}
where $\gamma'_{f\uparrow}\!=\!-i(f_{\uparrow}\!-\!f_{\uparrow}^{\dagger})$.
A (time-ordered) correlation function $G$ of a physical operator $O$ that is not invariant
under the particle-hole transformation $\mathcal{D}_{\uparrow}$ must
be calculated in the SS representation \emph{with} the projector,
\begin{equation}
G=2\ev{\hat{T}_{\tau}O_{SS}(\tau_{1})O_{SS}(\tau_{2})\mathbb{P}}_{SS},\label{eq:Pimp}
\end{equation}
where $O_{SS}$ is the SS representation of the physical operator
$O$. The factor of 2 is because $\mathcal{Z}\!=\!\mathcal{Z}_{SS}/2$. As the expectation value on the right-hand side (RHS) is 
taken with respect to the quadratic slave-spin Hamiltonian\emph{ $H_{SS}$, }Wick's theorem
can be used to explicitly implement $\mathbb{P}$ and calculate $G$ exactly.

\emph{Impurity edge-coupled to the Kitaev chain.}---As an application and
concrete demonstration of our results, we now specialize to the case
of an impurity hybridizing with the end of a semi-infinite Kitaev chain~\cite{Kitaev2001}. This special case is hereafter referred to as
the KMAI (Kitaev Majorana-Anderson impurity) model. The SS representation
of the KMAI model is obtained by using $H_{C}\!=\!H_{K}$ and $V_{i}\!=\!V\delta_{i1}$
in Eq. (\ref{eq:ssH2}), where
\begin{equation}
H_{K}=\sum_{j=1}^{\infty}\left[(-tc_{j}^{\dagger}c_{j+1}+\Delta c_{j}c_{j+1}+\mathrm{h.c.})-\mu c_{j}^{\dagger}c_{j}\right],\label{eq:Hk}
\end{equation}
 describes a semi-infinite Kitaev chain with hopping integral \emph{t},
\emph{p}-wave pairing amplitude $\Delta$, and chemical potential
$\mu.$ The physical Green's functions (GFs)
for $d_{\downarrow}\,(d_{\uparrow})$, calculable without (with)
constraint, are obtained in the SS representation as a product
of free-fermion imaginary-time slave GFs. For example, the $d_{\downarrow}$-fermion
GF is given by
\begin{equation}
\mathcal{G}_{d\downarrow}(\tau) =-\ev{\hat{T}_{\tau}\Gamma_{\uparrow}^{y}(\tau)\Gamma_{\uparrow}^{y}(0)\Gamma_{\uparrow}^{z}(\tau)\Gamma_{\uparrow}^{z}(0)f_{\downarrow}(\tau)f_{\downarrow}^{\dagger}(0)}_{SS},\label{eq:downGF}
\end{equation}
 where the RHS can be Wick contracted. In the Matsubara frequency domain, this becomes a convolution product, which after analytic continuation to real frequencies gives rise
to temperature $(T)$\emph{ }dependence in the spectral functions
of the physical impurity fermions $(d_{\sigma})$. This emphasizes
that the latter are interacting, even though the slave fermions are not.
The one-particle slave-fermion GFs appearing on the RHS of Eq. (\ref{eq:downGF}) after Wick contraction
can be calculated exactly using boundary GF methods~\cite{SuppMat}. When $\epsilon\!=\!0$, $H$ enjoys full particle-hole
(ph) symmetry and $\mathcal{G}_{d\downarrow}$ is \emph{T}-independent
and given by
\begin{equation}
\mathcal{G}_{d\downarrow}^{\mathrm{ph}}(ik_{n})=\frac{ik_{n}-2V^{2}g_{\gamma_{1}}(ik_{n})}{(ik_{n})^{2}-4U^{2}-2ik_{n}V^{2}g_{\gamma_{1}}(ik_{n})},\label{eq:downGF2}
\end{equation}
 where $g_{\gamma_{1}}(\tau)=-\langle\hat{T}_{\tau}\gamma_{1}(\tau)\gamma_{1}(0)\rangle$, with
$\gamma_{1}=c_{1}+c_{1}^{\dagger}$, is the boundary GF of the
semi-infinite Kitaev chain in the absence of an impurity. Away from
particle-hole symmetry, $\mathcal{G}_{d\downarrow}(ik_{n})$ can only be given an integral expression,
but the spectral function has a simple form, 
\begin{multline}
A_{d\downarrow}(\omega,T)=2[1-2n_{F}(\epsilon)]\{n_{B}(\epsilon)n_{F}(\omega-\epsilon)\\
+[n_{B}(\epsilon)+1][1-n_{F}(\omega-\epsilon)]\}A_{d\downarrow}^{\mathrm{ph}}(\omega-\epsilon),\label{eq:downSP}
\end{multline}
where $A_{d\downarrow}^{\mathrm{ph}}(\omega)$ is the $T$-independent, particle-hole symmetric
spectral function obtained from Eq. (\ref{eq:downGF2}), and $n_{B}$
($n_{F}$) is the Bose (Fermi) function. The first term in Eq.~(\ref{eq:downSP}) corresponds to the absorption of a spin-$\uparrow$ bosonic density fluctuation of energy $\epsilon$ by a spin-$\downarrow$ fermion of energy $\omega\!-\!\epsilon$, while the second term describes the emission, stimulated or spontaneous, of such a density fluctuation by a fermion of energy $\omega$. Turning now to the hybridizing $d_{\uparrow}$ impurity fermion, its Matsubara GF can be calculated by explicitly implementing the projector $\mathbb{P}$ using Eq.~(\ref{eq:projF}), which yields
\begin{equation}
\mathcal{G}_{d\uparrow}(ik_{n})=\frac{ik_{n}-V^{2}g_{\gamma_{1}}(ik_{n})+2U[2n_{F}(\epsilon)-1]}{(ik_{n})^{2}-4U^{2}-2ik_{n}V^{2}g_{\gamma_{1}}(ik_{n})}.\label{eq:upGF}
\end{equation}
An expression for $A_{d\uparrow}(\omega,T)$ can be obtained from the analytic continuation of Eq.~(\ref{eq:upGF}) to real frequencies.

\begin{figure}
\includegraphics[width=0.45\textwidth,height=0.4\textwidth,keepaspectratio]{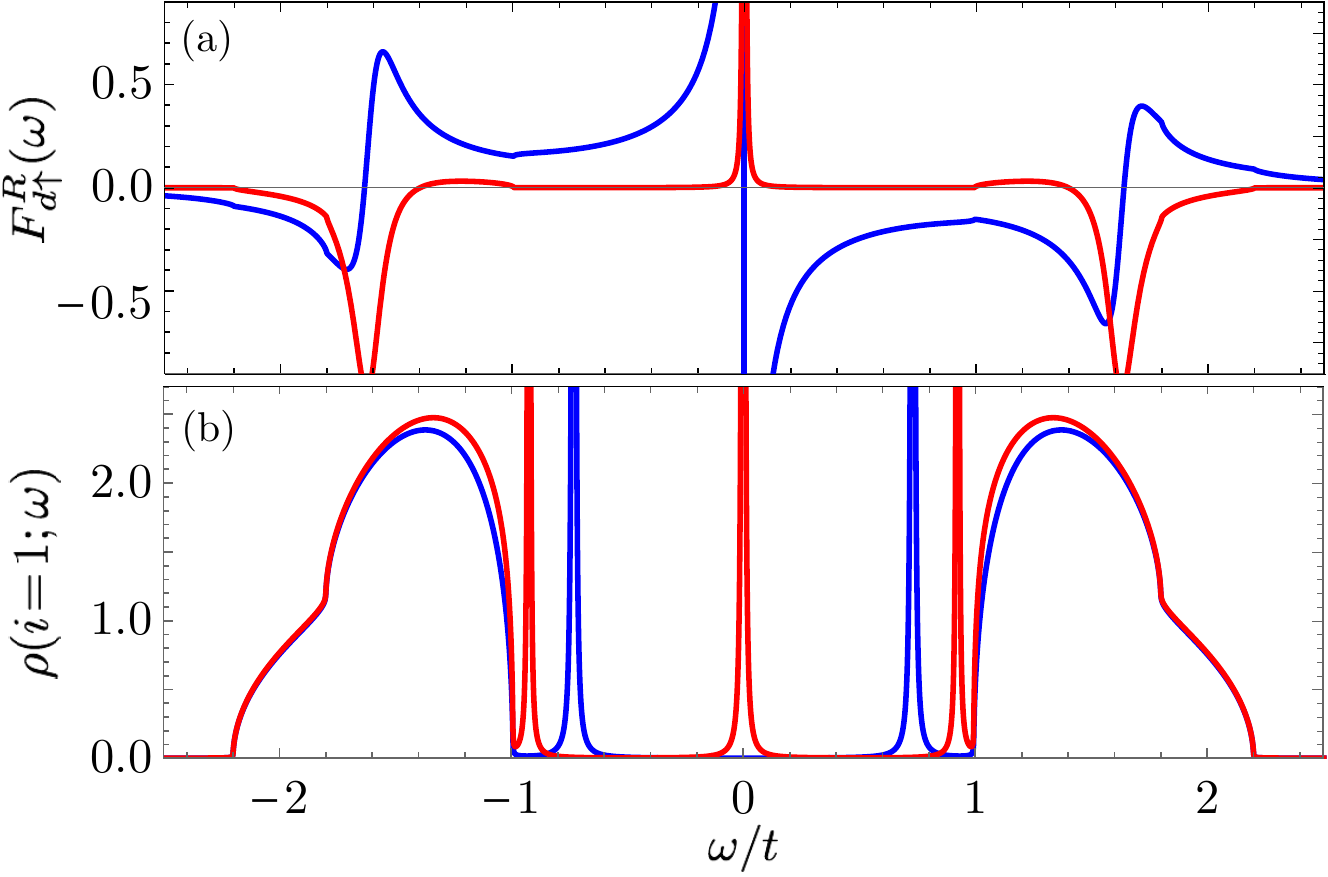}
\caption{(a) Real (blue) and imaginary (red) parts of the impurity retarded Gor'kov function $F_{d\uparrow}^{R}(\omega)$ for a Kitaev chain in the topological phase. Parameters are chosen as $\mu\!=\!0.2t,\Delta\!=\!0.5t,V\!=\!0.4t,U\!=\!0.7t.$ (b) Interaction dependence of the boundary density of states $\rho(i\!=\!1,\omega)$ of the $c$-fermions, for $\mu\!=\!0.2t,\Delta\!=\!0.5t,V\!=\!0.4t$, and $U\!=\!0$ (blue), $U\!=\!0.3t$ (red).}
\label{fig:oddfreq}
\end{figure}

\emph{Odd-frequency pairing.}---The Majorana hybridization with the
Kitaev chain results in proximity-induced superconductivity for the
$d_{\uparrow}$-fermions. The only possibility in this case is pure
odd-frequency pairing~\cite{Lee2017}, characterized by the real (imaginary) part
of the retarded Gor'kov function being odd (even) in frequency~\cite{berezinskii1974,kirkpatrick1991,balatsky1992} (Fig.~\ref{fig:oddfreq}a). The latter
is obtained by analytic continuation of the Matsubara Gor'kov function,
\begin{equation}
\mathcal{F}_{d\uparrow}(ik_{n})=\frac{V^{2}g_{\gamma_{1}}(ik_{n})}{(ik_{n})^{2}-4U^{2}-2ik_{n}V^{2}g_{\gamma_{1}}(ik_{n})},\label{eq:upGork}
\end{equation}
where $g_{\gamma_{1}}(ik_{n})$ is odd in $ik_{n}$ by virtue of being
a Majorana GF~\cite{asano2013,huang2015}. Odd-frequency pairing on the impurity is a consequence of the particle-hole symmetric form (\ref{eq:pHhyb}) of the hybridization term, and in fact obtains regardless of the specific host Hamiltonian $H_C$.

\begin{figure}
\includegraphics[width=0.48\textwidth,height=0.8\textwidth,keepaspectratio]{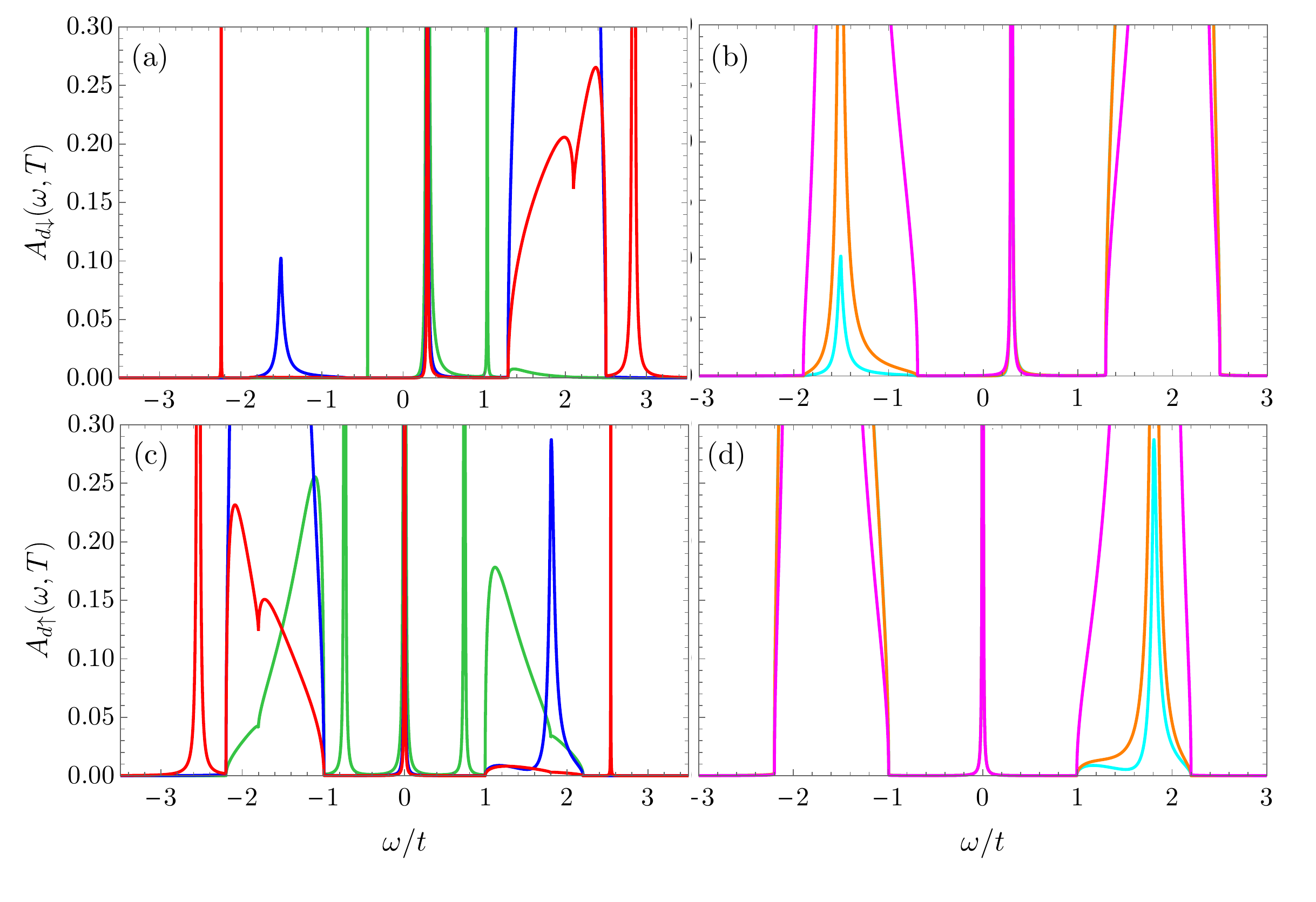}

\caption{(a)-(d) Spectral functions of $d_{\downarrow}$ (top row) and $d_{\uparrow}$
(bottom row) for various interaction strengths $U$ (left column)
and temperatures $T$ (right column), shown in the topological phase.
In all plots, $\mu\!=\!0.2t,\Delta\!=\!0.5t,V\!=\!0.4t,\epsilon\!=\!0.3t$ are fixed.
Left column: $T\!=\!0.05t$ and $U\!=\!0.05t$ (green), $U\!=\!0.8t$ (blue),
$U\!=\!1.2t$ (red). Right column: $U\!=\!0.8t$ and $T\!=\!0.05t$ (cyan), $T\!=\!0.07t$
(orange), $T\!=\!t$ (magenta).}
\label{fig:SFs}
\end{figure}

\emph{Impurity spectral functions.}---We now turn to the spectral functions $A_{d\sigma}(\omega)$ of the impurity fermions,
and restrict our discussion to the topological phase of the
KMAI model. The deviation $\epsilon$ from the particle-hole symmetric point sets the scale for the interaction-induced temperature dependence of those spectral functions. Low temperatures and $\epsilon\!>\!0$
accentuate the spectral asymmetry in $A_{d\downarrow}$ about
$\omega\!=\!\epsilon,$ shifting the spectral weight towards excitations
with $\omega\!>\!\epsilon$. It can be seen from Eq. (\ref{eq:downSP})
that, in the limit $T\!\gg\!\epsilon$, the temperature-dependent prefactors
tend towards unity, and particle-hole symmetry is restored (Fig.~\ref{fig:SFs}b).
This behavior with respect to temperature can be intuitively understood
in the atomic limit ($V\!=\!0$). In this limit, there are two infinitely sharp peaks in $A_{d\downarrow}$ at $\omega_{\pm}\!=\!\epsilon\!\pm\!2U$ corresponding
to localized charge excitations on the impurity. The spectral weight
for $\omega_{+}$ is greater as it is proportional to the $d_{\uparrow}$-fermion occupancy
$\ev{n_{d\uparrow}}$, which is favored over $d_{\downarrow}$-fermion occupancy
for $\epsilon\!>\!0$. Flipping the sign of $\epsilon$ reverses this
asymmetry, for $d_{\downarrow}$-fermion occupancy is then favored. This behavior
with respect to temperature carries over to the case when $V\!\neq\!0$. The temperature dependence of $A_{d\uparrow}$ can also be similarly
explained.

When the hybridization $V$ and interaction $U$ are both nonzero, both
impurity GFs have three poles (in the topological phase) which manifest as quasiparticle peaks in their spectral functions (Figs.~\ref{fig:SFs}a and \ref{fig:SFs}c). The two side peaks correspond
to impurity charge excitations, with a gap that increases monotonically
with $U$. For small $U$ and $V$, these excitations
feature as sharp peaks inside the energy
gap of the Kitaev SC. As $U$ or $V$ is increased, they fall into
the SC energy bands and broaden, and then eventually again become
sharp peaks when they move out of the bandwidth of the SC.
That the gap grows monotonically with $U$ is expected, as these states
differ in charge/occupancy.

The third quasiparticle peak (at $\omega\!=\!\epsilon$ for $A_{d\downarrow}$ and $\omega\!=\!0$ for $A_{d\uparrow}$) is never broadened and persists
for any non-zero $U,\,V$. We consider the $\omega\!=\!\epsilon$
peak in $A_{d\downarrow}$. This is where a sharp peak would
occur were the $d_{\downarrow}$ free $(U\!=\!0)$, but it is not and
the peak persists for large $U$. This is an indirect signature of the presence
of a MZM, as can be understood from the small $U/V$ limit. A semi-infinite
Kitaev chain implies there must be an exact MZM at zero energy.
But the original MZM $(c_{1}\!+\!c_{1}^{\dagger})$ of the Kitaev chain
is now paired with $d_{\uparrow}\!+\!d_{\uparrow}^{\dagger}$ to form
a local complex fermion due to $H_{{\rm hyb}}$. Neither of the two
Majorana modes that make up $d_{\downarrow}$ can be the new MZM as
$n_{d\downarrow}$ is conserved. Therefore, $-i(d_{\uparrow}\!-\!d_{\uparrow}^{\dagger})$
must be the new MZM in the small-$U/V$ limit. As it has to be an
exact zero mode, interactions cannot change its energy. In this limit,
the $d_{\downarrow}$ becomes free, and this features as a sharp
peak in $A_{d\downarrow}$ at $\omega=\epsilon.$ That $-i(d_{\uparrow}\!-\!d_{\uparrow}^{\dagger})$
is the preferred MZM in this limit features as a sharp peak
at $\omega=0$ in $A_{d\uparrow}.$ In the opposite large-$U/V$ limit, energetics suggest that the
original mode $(c_{1}\!+\!c_{1}^{\dagger})$ is the preferred MZM.

An obvious check of this intuitive reasoning is provided by the $c$-fermion local density of states (LDOS) at the boundary---there must be an MZM peak at any finite $U$,
with spectral weight that \emph{increases }with $U$. The local GFs
for the $c$-fermions can be calculated on an arbitrary lattice site~\cite{SuppMat}, from which the
corresponding LDOS can be obtained. The boundary LDOS (Fig.~\ref{fig:oddfreq}b) supports our intuition: an MZM peak appears for
any nonzero interaction and its spectral weight obtained by numerical
integration does increase with $U$. The two other subgap states
are non-topological Andreev bound states induced by the impurity, reminiscent of Yu-Shiba-Rusinov states~\cite{yu1965,shiba1968,rusinov1969}.

\begin{figure}
\includegraphics[width=0.45\textwidth,height=0.4\textwidth,keepaspectratio]{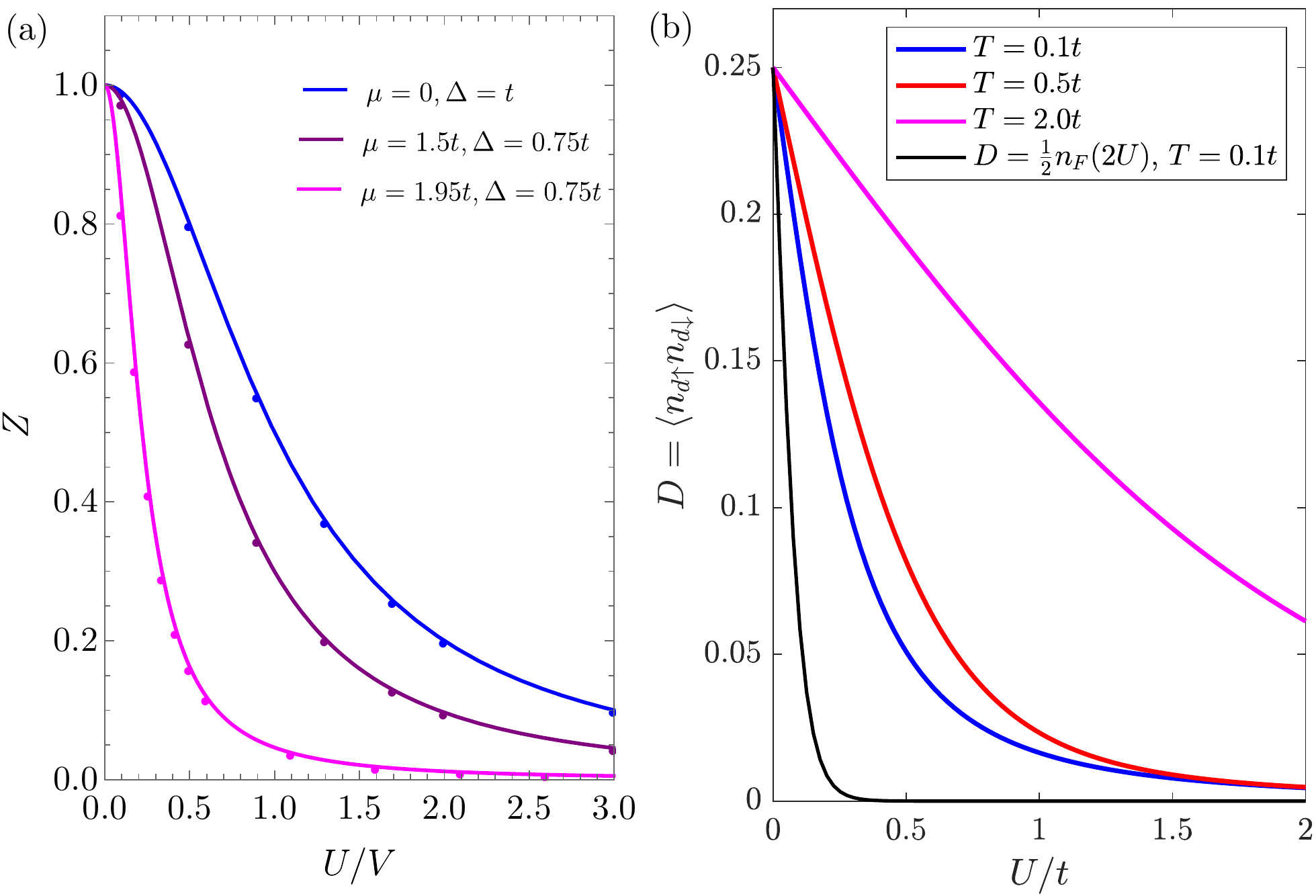}
\caption{(a) Interaction dependence of the $d_\downarrow$-fermion quasiparticle weight $Z$, for several values of $\mu$ and $\Delta$, which control the localization length of the original end MZM in the Kitaev chain. Continuous curves correspond to Eq.~(\ref{eq:Z}), while dots are the result of numerically integrating $A_{d\downarrow}(\omega,T$) over
a small neighborhood of $\omega\!=\!\epsilon$. (b) Interaction
dependence of impurity double occupancy $D$ for various $T$ in the particle-hole symmetric limit ($\epsilon\!=\!0$). Black: atomic limit ($V\!=\!0$), all other curves: $\mu\!=\!0.2t,\Delta\!=\!0.5t,V\!=\!0.4t,\epsilon\!=\!0$.}
\label{fig:ZandD}
\end{figure}

\emph{Local Fermi liquid.}---Since the free-fermion peak in $A_{d\downarrow}$ remains sharp even in the presence of interactions, a natural quantity to study is the associated quasiparticle weight $Z$. This can be calculated from Eqs. (\ref{eq:downGF2})-(\ref{eq:downSP})
and is given by
\begin{equation}
Z=\frac{1}{1+(2/\lambda)(U/V)^{2}},\label{eq:Z}
\end{equation}
 where $\lambda(\mu,\Delta)$ is the spectral weight (characterizing
the localization) of the MZM peak in the boundary LDOS of a noninteracting Kitaev
chain with no impurity~\cite{SuppMat}. In the noninteracting limit, the $d_{\downarrow}$-fermion is
free and so $Z\!=\!1$. The interaction renormalizes $Z$ to a value less
than one (Fig.~\ref{fig:ZandD}a), and transfers some spectral
weight to other excitations, thus giving credence to a \emph{local
Fermi liquid} picture for the $d_{\downarrow}$-fermion. This holds only
in the topological phase, as the free-fermion peak for finite $U$
and $V$ has its origins in $-i(d_{\uparrow}\!-\!d_{\uparrow}^{\dagger})$
being an MZM candidate, which is not true in the trivial phase. It
is also not valid for the hybridizing $d_{\uparrow}$-fermion, as
the spectral weight of the $\omega\!=\!0$ peak is trivially less than one due to proximity coupling with the
Kitaev chain, even in the absence of interactions. Also, conforming with the intuitive discussion in the
previous section, $Z$ is suppressed at large $U$, when $c_{1}\!+\!c_{1}^{\dagger}$
is the preferred MZM. 

Another measure of interparticle correlations on the QD is provided by the mean squared density fluctuation $D\!=\!(1/2)\langle(n_d\!-\!\ev{n_d})^{2}\rangle,$ where $n_d\!=\!n_{d\uparrow}\!+\!n_{d\downarrow}$.
In the particle-hole symmetric limit ($\epsilon\!=\!0$), because $\ev{n_d}\!=\!1$ this reduces to the double occupancy $D\!=\!\ev{n_{d\uparrow}n_{d\downarrow}}$,
which can be calculated from a derivative of the logarithm of the partition function with respect to $U$, to get 
\begin{equation}
D=\frac{1}{4}+\frac{U}{2}\int\frac{d\omega}{2\pi}A_{(d+d^{\dagger})\uparrow}(\omega)\frac{n_{F}(\omega)}{\omega},\label{eq:D}
\end{equation}
 where $A_{(d+d^{\dagger})\uparrow}(\omega)$ is the spectral function
of the hybridizing Majorana mode $d_{\uparrow}\!+\!d_{\uparrow}^{\dagger}$. The Matsubara GF of this operator is simply the sum of electron, hole, and Gor'kov GFs
of the $d_{\uparrow}$-fermion. Plots of $D$ (Fig.~\ref{fig:ZandD}b)
reveal that density fluctuations are suppressed at large $U$ and
low $T$, but encouraged by hybridization $V$.

\emph{Departures from exact solvability.}---We now consider deviations from the exactly solvable point $\epsilon\!=\!h$. Defining $\delta\!=\!(\epsilon\!-\!h)/2$, the SS Hamiltonian (\ref{eq:ssH2}) becomes
\begin{equation}\label{HSSprimed}
H_{SS}\!=\!H_{SS}(\epsilon\!=\!h)-\delta(n_{f\downarrow}\!-\!1/2)-i\delta\Gamma_{\uparrow}^{x}\Gamma_{\uparrow}^{y}(n_{f\downarrow}\!-\!1/2),
\end{equation}
 where $H_{SS}(\epsilon\!=\!h)$ is the bilinear exactly solvable part. For sufficiently small $\delta$, corrections to physical observables away from the exactly solvable limit can be computed by treating the last term in Eq.~(\ref{HSSprimed}) in perturbation theory, in analogy to the perturbative analysis of small departures from the Toulouse point in the Kondo problem~\cite{wiegmann1978}. We emphasize this is distinct from ordinary perturbation theory in the physical interaction strength $U$; here $U$ can be arbitrarily large, and the perturbation corresponds to either a shift
in the chemical potential of the impurity fermions or a change in the Zeeman field. For example,
to linear order in $\delta$, the free energy is $F=F^{(0)}+F^{(1)}\delta+\mathcal{O}(\delta^2)$ where
\begin{equation}
F^{(1)}=2[1-2 n_F(\epsilon)][1/4-D]-n_F(\epsilon),
\end{equation}
with $D$ the $T$-dependent double occupancy in the particle-hole symmetric limit, given in Eq.~(\ref{eq:D}).

\emph{Outlook.}---Several extensions of our work are possible. Besides different choices of bath Hamiltonian, such as 2D or 3D topological SCs or Majorana hopping models, our exact solution trivially generalizes to periodic Majorana-Anderson models, where the impurity fermions acquire a lattice-site index. However, the $\mathbb{Z}_2$ slave-spin solution of such models involves a local projection on every site, as in the Majorana-Falicov-Kimball model~\cite{Prosko2017}, which likely limits exact sovability to the computation of correlation functions of operators that  commute with the local particle-hole transformation $\mathcal{D}_{\uparrow}$ [see Eq.~(\ref{eq:Pimp})]. While applications to spin-polarized topological SCs naturally justify a spin-selective choice (\ref{eq:pHhyb}) of hybridization term~\cite{hoffman2017,lee2013,Cheng2014,liu2015}, it is also possible to generalize the latter such that multiple Majorana
modes on the QD hybridize equally with the bath fermions while retaining exact solvability.

\emph{Acknowledgements.}---We thank R. Boyack for helpful discussions. JM was supported by NSERC grant \#RGPIN-2014-4608, the CRC Program, CIFAR, and the University of Alberta.

\bibliographystyle{apsrev4-1}
\bibliography{MAIREFS}

\end{document}


\renewcommand{\thetable}{S\Roman{table}}
\renewcommand{\thefigure}{S\arabic{figure}}
\renewcommand{\thesection}{S\Roman{section}}
\renewcommand{\thesubsection}{S\arabic{subsection}}
\renewcommand{\theequation}{S\arabic{equation}}

\title{Supplemental Material for ``Exactly solvable Majorana-Anderson impurity models''}

\author{G. Shankar}
\affiliation{Department of Physics, University of Alberta, Edmonton, Alberta T6G 2E1, Canada}
\author{J. Maciejko}
\affiliation{Department of Physics, University of Alberta, Edmonton, Alberta T6G 2E1, Canada}
\affiliation{Theoretical Physics Institute, University of Alberta, Edmonton, Alberta T6G 2E1, Canada}
\affiliation{Canadian Institute for Advanced Research, Toronto, Ontario M5G 1Z8, Canada}

\date\today
\maketitle

This supplemental material provides a detailed proof of the disappearance of the gauge constraint in the $\mathbb{Z}_2$ slave-spin formulation of the Majorana-Anderson impurity (MAI) model (Sec.~\ref{sec:constraint}), as well as details of the calculation of boundary Green's functions in the Kitaev Majorana-Anderson impurity (KMAI) model (Sec.~\ref{sec:boundaryGF}).

\section{Correlation functions without constraint}
\label{sec:constraint}

The following proof is adapted from Ref.~\cite{Prosko2017}. Let $G$
be a correlation function of $M$ operators $O_{1},...,O_{M}$ constructed
out of the physical fermion operators $\left\{ c,c^{\dagger},d_{\uparrow},d_{\uparrow}^{\dagger},d_{\downarrow},d_{\downarrow}^{\dagger}\right\} $
such that $\forall i,\,\left[O_{i},\mathcal{D}_{\uparrow}\right]\!=\!0$,
where $\mathcal{D}_{\uparrow}d_{\uparrow}\mathcal{D}_{\uparrow}^{-1}\!=\! d_{\uparrow}^{\dagger}$.
Considering a given imaginary-time ordering of the operators $\{O_{i}\}$,
we have 
\begin{align}
G  =\ev{O_{1}(\tau_{1})\cdots O_{M}(\tau_{M})} =\frac{1}{\mathcal{Z}}\tr e^{-\beta H}\prod_{i=1}^{M}e^{\tau_{i}H}O_{i}e^{-\tau_{i}H}.
\end{align}
Inserting two identity operators inside the trace as $\mathcal{D_{\uparrow}}\mathcal{D_{\uparrow}}^{-1}$
and then using the cyclicity of the trace,
\begin{align}
G & =\frac{1}{\mathcal{Z}(V,U,\epsilon,h)}\tr\mathcal{D_{\uparrow}}^{-1}\mathcal{D_{\uparrow}}e^{-\beta H(V,U,\epsilon,h)}\mathcal{D_{\uparrow}}^{-1}\mathcal{D_{\uparrow}}\prod_{i=1}^{M}e^{\tau_{i}H(V,U,\epsilon,h)}O_{i}e^{-\tau_{i}H(V,U,\epsilon,h)}\nonumber \\
 & =\frac{1}{\mathcal{Z}(V,U,\epsilon,h)}\tr e^{-\beta H(V,-U,h,\epsilon)}\mathcal{D_{\uparrow}}\left[\prod_{i=1}^{M}e^{\tau_{i}H(V,U,\epsilon,h)}O_{i}e^{-\tau_{i}H(V,U,\epsilon,h)}\right]\mathcal{D}_{\uparrow}^{-1},\label{eq:Gcnstrt1}
\end{align}
 where in the last step, the result $\mathcal{D}_{\uparrow}H(V,U,\epsilon,h)\mathcal{\mathcal{D}}_{\uparrow}^{-1} \!=\! H(V,-U,h,\epsilon)$
from the main text has been used. Again inserting the identity operator
$\mathcal{D_{\uparrow}}\mathcal{D_{\uparrow}}^{-1}$ multiple times inside the product,
\begin{align}
G & =\frac{1}{\mathcal{Z}(V,U,\epsilon,h)}\tr e^{-\beta H(V,-U,h,\epsilon)}\left[\prod_{i=1}^{M}e^{\tau_{i}H(V,-U,h,\epsilon)}\mathcal{D}_{\uparrow}O_{i}\mathcal{D}_{\uparrow}^{-1}e^{-\tau_{i}H(V,-U,h,\epsilon)}\right]\nonumber \\
 & =\frac{1}{\mathcal{Z}(V,-U,h,\epsilon)}\tr e^{-\beta H(V,-U,h,\epsilon)}\left[\prod_{i=1}^{M}e^{\tau_{i}H(V,-U,h,\epsilon)}O_{i}e^{-\tau_{i}H(V,-U,h,\epsilon)}\right],\label{eq:Gcnstrt2}
\end{align}
where the second step follows from the assumption: $\forall i,\,\left[O_{i},\mathcal{D}_{\uparrow}\right]\!=\!0$,
and from $\mathcal{Z}(V,U,\epsilon,h)\!=\!\mathcal{Z}(V,-U,h,\epsilon)$, as shown in the
main text. We therefore have the result $G(V,U,\epsilon,h)\!=\!G(V,-U,h,\epsilon)$.
In the slave-spin (SS) representation, the same correlation function
$G$ is represented as
\begin{equation}
G=\frac{2}{\mathcal{Z}_{SS}}\tr e^{-\beta H}\left[\prod_{i=1}^{M}e^{\tau_{i}H_{SS}}O_{i}^{(SS)}e^{-\tau_{i}H_{SS}}\right]\mathbb{P},
\end{equation}
where the operators $\{O_{i}^{(SS)}\}$ are now constructed out of
fermion operators $\left\{ c,c^{\dagger},f_{\uparrow},f_{\uparrow}^{\dagger},f_{\downarrow},f_{\downarrow}^{\dagger}\right\} $
that act on the slave-spin Hilbert space. The factor of 2 arises because $\mathcal{Z}\!=\!\mathcal{Z}_{SS}/2$, as discussed in the main text. Since the action of $\mathcal{D}_{\uparrow}$
is implemented in the slave-spin representation by $\mu^{x}$, repeating
the derivations in Eqs. (\ref{eq:Gcnstrt1})-(\ref{eq:Gcnstrt2})
in the SS representation, making use of the result $\mu^{x}\mathbb{P}\mu^{x}\!=\!1-\mathbb{P}$,
yields the result 
\begin{equation}
G=\frac{1}{\mathcal{Z}_{SS}}\tr e^{-\beta H_{SS}}\left[\prod_{i=1}^{M}e^{\tau_{i}H_{SS}}O_{i}^{(SS)}e^{-\tau_{i}H_{SS}}\right].
\end{equation}
 Therefore, correlation functions of operators that commute with the
operator $\mathcal{D}_{\uparrow}$ can be calculated in the SS representation
without constraint.

\section{Boundary and local Green's functions in the KMAI model}
\label{sec:boundaryGF}

Defining a new complex slave-fermion $\eta_{\uparrow}\!=\!(\Gamma_{\uparrow}^{y}\!+\!i\Gamma_{\uparrow}^{x})/2$,
the exactly solvable SS representation of the KMAI model can be rewritten
as
\begin{align}
H_{SS}= & \sum_{i=1}^{\infty}\left[-tc_{i}^{\dagger}c_{i+1}+\Delta c_{i}c_{i+1}+\text{h.c.}\right]-\mu\sum_{i=1}^{\infty}c_{i}^{\dagger}c_{i}\nonumber \\
 & -V\left(c_{1}\eta_{\uparrow}+c_{1}^{\dagger}\eta_{\uparrow}+\text{h.c.}\right)+2U\eta_{\uparrow}^{\dagger}\eta_{\uparrow}+\epsilon(f_{\downarrow}^{\dagger}f_{\downarrow}-1/2).\label{eq:KMAIF}
\end{align}
We first calculate the boundary Green's functions (GFs) of the slave-fermions
and then discuss how the physical impurity $(d_{\sigma})$ GFs can
be obtained from these. We may write $H_{SS}$ in Bogoliubov-de-Gennes
(BdG) form with Nambu spinor $\Psi\!=\!(f_{\downarrow}\quad f_{\downarrow}^{\dagger}\quad\eta_{\uparrow}\quad\eta_{\uparrow}^{\dagger}\quad c_{1}\quad c_{1}^{\dagger}\quad c_{2}\quad c_{2}^{\dagger}\quad\cdots)^\intercal$
and BdG matrix $h_{BdG}$ as 
\begin{equation}
H_{SS}=\frac{1}{2}\Psi^{\dagger}\left(\begin{array}{cc|c|ccc}
\epsilon\sigma^{z} & 0 & 0 & 0 & 0 & \dots\\
0 & 2U\sigma^{z} & C & 0 & 0 & \dots\\
\hline 0 & C^{\dagger} & -\mu\sigma^{z} & T & 0 & \dots\\
\hline 0 & 0 & T^{\dagger} & -\mu\sigma^{z} & T & 0\\
0 & 0 & 0 & T^{\dagger} & -\mu\sigma^{z} & T\\
\vdots & \vdots & \vdots & 0 & T^{\dagger} & \ddots
\end{array}\right)\Psi,\label{eq:bdg}
\end{equation}
 where the matrices $T$ and $C$ are defined as 
\begin{equation}
T=\left(\begin{array}{cc}
-t & -\Delta\\
\Delta & t
\end{array}\right),\quad C=\left(\begin{array}{cc}
-V & -V\\
V & V
\end{array}\right).
\end{equation}
Partitioning the resolvent matrix $\mathbf{G}\!=\!(z-h_{BdG})^{-1}$ in
correspondence with the partitions of $h_{BdG}$, we write 
\begin{equation}
\mathbf{G}=\left(\begin{array}{ccc}
\mathbf{G}_{A} & \mathbf{G}_{AS} & \mathbf{G}_{AB}\\
\mathbf{G}_{SA} & \mathbf{G}_{S} & \mathbf{G}_{SB}\\
\mathbf{G}_{BA} & \mathbf{G}_{BS} & \mathbf{G}_{B}
\end{array}\right).
\end{equation}
The GFs of the $\eta_\uparrow$ and $f_{\downarrow}$-fermions are obtained
from $\mathbf{G}_{A}$, which is the part of $\mathbf{G}$ that corresponds
to the first diagonal partition of $h_{BdG}$, and represents the
Anderson impurity. Solving for $\mathbf{G}_{A}$ from $(z-h_{BdG})\mathbf{G}\!=\!\mathbb{I}$,
we obtain
\begin{equation}
\mathbf{G}_{A}^{-1}=\left(\begin{array}{cc}
z-\epsilon\sigma^{z} & 0\\
0 & z-2U\sigma^{z}-V^{2}\left[\sum_{\alpha,\beta=1}^{2}\mathbf{g}_{S}^{\alpha\beta}(z)\right]\left(1-\sigma^{x}\right)
\end{array}\right),\label{eq:GAinv}
\end{equation}
 where \textbf{$\mathbf{g}_{S}(z)$ }is the left boundary (Nambu)
GF of the Kitaev chain (without an impurity). 	The sum of all matrix elements of $\mathbf{g}_{S}(z)$ is simply
the frequency representation of the Majorana GF $g_{\gamma_{1}}(\tau)\!=\!-\ev{T_{\tau}\gamma_{1}(\tau)\gamma_{1}(0)}$, where $\gamma_{1}\!=\!c_{1}\!+\!c_{1}^{\dagger}$. An explicit expression
for $\mathbf{g}_{S}(z)$ is obtained following the method outlined
in Appendix A of Ref.~\cite{Jonckeere2017}. Since the result quoted
there contains typos, we state here the corrected result in their
notation:
\begin{equation}
\mathbf{g}_{S}(z)=\mathbf{g}_{B}(0;z)-\mathbf{g}_{B}(1-0;z)\mathbf{g}_{B}^{-1}(0;z)\mathbf{g}_{B}(0-1;z),
\end{equation}
where
\begin{align}
\mathbf{g}_{B}(0;z) & =\left(z-\mu\sigma^{z}\right)\mathcal{F}_{-1}(z)+2t\sigma^{z}\mathcal{F}_{0}(z)\,\,,\\
\mathbf{g}_{B}(1-0;z) & =2i\Delta\mathcal{F}_{-1}(z)\sigma^{y}-\left(z-\mu\sigma^{z}\right)\mathcal{F}_{0}(z)+\left(2t\sigma^{z}+2i\Delta\sigma^{y}\right)\left[\frac{1}{4\left(t^{2}-\Delta^{2}\right)}-\mathcal{F}_{1}(z)\right]\,\,,\\
\mathbf{g}_{B}(0-1;z) & =-2i\Delta\mathcal{F}_{-1}(z)\sigma^{y}-\left(z-\mu\sigma^{z}\right)\mathcal{F}_{0}(z)+\left(2t\sigma^{z}-2i\Delta\sigma^{y}\right)\left[\frac{1}{4\left(t^{2}-\Delta^{2}\right)}-\mathcal{F}_{1}(z)\right]\,\,,
\end{align}
 and the functions $\mathcal{F}_{m}(z)$ with $m\in\{-1,0,1\}$ are
given by
\begin{align}
\mathcal{F}_{m}(z) & =\frac{1}{4\left(t^{2}-\Delta^{2}\right)}\cdot\frac{1}{Q_{+}(z)-Q_{-}(z)}\sum_{s=\pm1}\frac{sQ_{s}^{m}(z)}{\sqrt{1-\nicefrac{1}{Q_{s}^{2}(z)}}},\\
\mathrm{where}\quad Q_{\pm}(z) & =\frac{1}{2\left(\Delta^{2}-t^{2}\right)}\left[-t\mu\pm\sqrt{\Delta^{2}\mu^{2}-\left(\Delta^{2}-t^{2}\right)\left(z^{2}-4\Delta^{2}\right)}\right].
\end{align}
Inside the superconducting gap, the retarded GF of the MZM $(\gamma_1)$ of a Kitaev chain takes the form $g_{\gamma_1}(\omega)\!=\!\lambda(\mu,\Delta)/(\omega+i\eta)$, where $\eta$ is a positive infinitesimal. For example, $\lambda(0,t)\!=\!2$ and thus $g_{\gamma_1}(\omega)$ is a free Majorana GF, which reflects the fact that the MZM is exactly localized at the boundary and decoupled from the bulk. Away from this special point in the phase diagram of the Kitaev chain, $\lambda\!<\!2$ as the localization length of the MZM is finite.

As shown in the main text, all physical impurity Matsubara GFs can be obtained
as convolutions of GFs of slave-fermion operators $\{\Gamma_{\uparrow}^{x},\Gamma_{\uparrow}^{y},\Gamma_{\uparrow}^{z},f_{\downarrow},\gamma'_{f\uparrow}\!=\!-i(f_{\uparrow}\!-\!f_{\uparrow}^{\dagger})\}$.
From Eq. (\ref{eq:KMAIF}), it is easy to see that the GFs of $\Gamma_{\uparrow}^{z}$,
$f_{\downarrow}$, and $\gamma'_{f\uparrow}$ respectively are
\begin{align}
\mathcal{G}_{\uparrow}^{zz}(ik_{n}) & =\frac{2}{ik_{n}},\\
\mathcal{G}_{f\downarrow}(ik_{n}) & =\frac{1}{ik_{n}-\epsilon},\\
\mathcal{G}_{\gamma'_{f\uparrow}}(ik_{n}) & =\frac{2}{ik_{n}}.
\end{align}
The factor of 2 in the Majorana GFs is because Majorana operators
satisfy the Clifford algebra: for example, $\acomm{\Gamma_{\uparrow}^{\alpha}}{\Gamma_{\uparrow}^{\beta}}=2\delta_{\alpha\beta}$.
Since $\eta_{\uparrow} \!=\! \Gamma_{\uparrow}^{x} \!+\!i \Gamma_{\uparrow}^{y}$,
the GFs of $\Gamma_{\uparrow}^{x}$ and $\Gamma_{\uparrow}^{y}$ can
be calculated from appropriate linear combinations of the matrix elements
of the Nambu GF of $(\eta_{\uparrow}\quad\eta_{\uparrow}^{\dagger})^\intercal$,
the inverse of which is given by the second diagonal block of Eq.
(\ref{eq:GAinv}).

The local GFs of the host $c_{j}$-fermions can also be calculated
in the same framework, by repartitioning and considering appropriate
blocks of the resolvent matrix $\mathbf{G}$. Let $\mathbf{G}_{c}(j;z)$
denote the Nambu GF of $(c_{j}\quad c_{j}^{\dagger})^\intercal$. Since
$\mathbf{G}_{c}(j\!=\!1;z)\!=\!\mathbf{G}_{S}(z)$, solving $(z-h_{BdG})\mathbf{G}\!=\!\mathbb{I}$
for $\mathbf{G}_{S}$ yields 
\begin{equation}
\mathbf{G}_{c}^{-1}(j=1;z)=\mathbf{g}_{S}^{-1}(z)-\frac{2V^{2}z}{z^{2}-4U^{2}}\left(1+\sigma^{x}\right).
\end{equation}

To calculate $\mathbf{G}_{c}(j\!>\!1;z)$, we make use of the Dyson equation
\begin{equation}
\mathbf{G}_{c}^{-1}(j>1;z)=\mathbf{g}_{S}^{-1}(z)-T^{\dagger}\rho_{j-1}(j-1;z)T,
\end{equation}
 where $\rho_{j\!-\!1}(j\!-\!1;z)$ is the\emph{ right boundary} Nambu GF
of a \emph{finite $(j\!-\!1)$}-site Kitaev chain coupled to an Anderson
impurity at the \emph{left boundary}. This is a finite KMAI system
with a finite BdG matrix $h_{BdG}^{(j\!-\!1)}$ that is obtained simply
by truncating $h_{BdG}$ in Eq. (\ref{eq:bdg}) appropriately (at
the $(j\!-\!1)$-th $\mu\sigma^{z}$ block). The Dyson equation for $\rho_{j\!-\!1}(j\!-\!1;z)$
reads 
\begin{align}
\rho_{j-1}(j-1;z) & =\left[z+\mu\sigma^{z}-T^{\dagger}\rho_{j-2}(j-2;z)T\right]^{-1}\nonumber \\
 & =T^{-1}\left[(z+\mu\sigma^{z})T^{-1}-T^{\dagger}\rho_{j-2}(j-2;z)\right]^{-1}.
\end{align}
 The right-hand side of this equation is a matrix Möbius transformation
\footnote{Given $M\times M$ matrices $\mathbf{a},\mathbf{b},\mathbf{c},\mathbf{d}$
and $\mathbf{x}$, the matrix M{\"o}bius transformation of $\mathbf{x}$ by $\Lambda$ is defined
as $\Lambda\bullet\mathbf{x}=\left(\mathbf{a}\mathbf{x}+\mathbf{b}\right)\left(\mathbf{c}\mathbf{x}+\mathbf{d}\right)^{-1}$
where $\Lambda=\left(\begin{array}{cc}
\mathbf{a} & \mathbf{b}\\
\mathbf{c} & \mathbf{d}
\end{array}\right)$ is a $2M\times2M$ matrix~\cite{Umerski1997}.} of $\rho_{j\!-\!2}(j\!-\!2;z)$,  that is,
\begin{align}
\rho_{j-1}(j-1;z)= & \left(\begin{array}{cc}
0 & T^{-1}\\
-T^{\dagger} & \left(z+\mu\sigma^{z}\right)T^{-1}
\end{array}\right)\bullet\rho_{j-2}(j-2;z)\nonumber \\
= & \left(\begin{array}{cc}
0 & T^{-1}\\
-T^{\dagger} & \left(z+\mu\sigma^{z}\right)T^{-1}
\end{array}\right)^{j-2}\bullet\rho_{1}(j=1;z),
\end{align}
 where $\rho_{1}(j\!=\!1;z)$ is the GF of a single site coupled to an
Anderson impurity, and so given by
\begin{align}
\rho_{1}(j=1;z)= & \left[z+\mu\sigma^{z}-\frac{2V^{2}z}{z^{2}-4U^{2}}(1+\sigma^{x})\right]^{-1}.
\end{align}

\bibliographystyle{apsrev4-1}
\bibliography{MAIREFS}